\documentclass[letterpaper]{jpconf}
\usepackage{graphicx}
\begin{document}
\title{Measurement of ${\cal B}(\Upsilon(nS) \to \mu^+\mu^-)$ at CLEO}

\author{Istv\'an Dank\'o (representing the CLEO Collaboration)}

\address{Rensselaer Polytechnic Institute, 110 8th St., Troy, New York 12180, USA}

\ead{idanko@mail.lns.cornell.edu}

\begin{abstract}
The decay branching fractions of the three narrow $\Upsilon$ resonances to $\mu^+ \mu^-$ have been measured by analyzing about 4.3 fb$^{-1}$ $e^+e^-$ data collected with the CLEO III detector. The branching fraction ${\cal B}(\Upsilon(1S) \to \mu^+ \mu^-) = (2.49 \pm 0.02 \pm 0.07)\%$ is consistent with the current world average but ${\cal B}(\Upsilon(2S) \to \mu^+ \mu^-) = (2.03 \pm 0.03 \pm 0.08)\%$ and ${\cal B}(\Upsilon(3S) \to \mu^+ \mu^-) = (2.39 \pm 0.07 \pm 0.10)\%$ are significantly larger than prior results. These new muonic branching fractions imply a narrower total decay width for the $\Upsilon$(2S) and $\Upsilon$(3S) resonances and lower other branching fractions that rely on these decays in their determination.
\end{abstract}

\section{Introduction}

Since their discovery, the long-lived $b\bar{b}$ states have played an essential role in the study of the strong interaction (QCD) \cite{Yreview}, and more recently they have become important in establishing the accuracy of Lattice QCD calculations \cite{LQCD}.
The large data samples collected recently by the CLEO detector in the vicinity of the $\Upsilon$(nS) ($n=1, 2, 3$) resonances enable us to determine the fundamental $b\bar{b}$ resonance parameters, such as the leptonic and total decay widths, with unprecedented precision. 

Since the total decay widths ($\Gamma$) of the $\Upsilon$ resonances below the open-beauty threshold are too narrow to be measured directly, one has to combine the leptonic branching fraction (${\cal B}_{\ell\ell}$) with the leptonic decay width ($\Gamma_{\ell\ell}$) in order to determine $\Gamma$: $\Gamma = \Gamma_{\ell\ell}/{\cal B}_{\ell\ell}$ \cite{Yreview,PDG}.
In practice, assuming lepton universality ($\Gamma_{ee} = \Gamma_{\mu\mu} = \Gamma_{\tau\tau}$), the leptonic decay width is replaced by $\Gamma_{ee}$, which can be extracted from the energy-integrated resonant hadron production cross section in $e^+e^-$ collisions, while the leptonic branching fraction is replaced by the muonic branching fraction, ${\cal B}_{\mu\mu} \equiv {\cal B}(\Upsilon \to \mu^+\mu^-$), which can be measured more accurately than ${\cal B}_{ee}$ or ${\cal B}_{\tau\tau}$. 

The leptonic branching fraction is also interesting in its own right since it represents the strength of the $\Upsilon$ decay to lepton pairs via annihilation to a virtual photon.
Furthermore, ${\cal B}_{\mu\mu}$ is generally used in determinations of the branching fractions of hadronic and electromagnetic transitions among the $\Upsilon$ states since these decays are often measured by observing the decay of the lower lying resonance to lepton pairs.
In addition, comparing ${\cal B}_{\mu\mu}$ to ${\cal B}_{ee}$ as well as to ${\cal B}_{\tau\tau}$ can provide a check of lepton universality and a test of new physics scenarios \cite{Higgs}. 

Based on previous measurements, ${\cal B}_{\mu\mu}$ has been established with a 2.4\% accuracy for the $\Upsilon$(1S) \cite{PDG}, and a modest 16\% and 9\% accuracy for the $\Upsilon$(2S) and $\Upsilon$(3S) \cite{early_exp}, respectively.
In this paper, we report the results of a recent measurement of ${\cal B}_{\mu\mu}$ for all three narrow $\Upsilon$ resonances by the CLEO collaboration \cite{CLEO_new}.

\section{Analysis technique}

The results are based on $1.1-1.2$ fb $^{-1}$ data collected on the peak of each resonance (``on-resonance samples'') as well as $0.2-0.4$ fb$^{-1}$ data collected $20-30$ MeV below each resonance (``off-resonance samples'').
The data were recorded by the CLEO~III detector at the Cornell Electron Storage Ring which is a symmetric $e^+ e^-$ collider.

To extract ${\cal B}_{\mu\mu}$, we measure the relative decay rate to muons and hadrons $\tilde{\cal{B}}_{\mu\mu} \equiv \Gamma_{\mu\mu}/\Gamma_{\rm had}=(\tilde{N}_{\mu\mu}/\varepsilon_{\mu\mu})/(\tilde{N}_{\rm had}/\varepsilon_{\rm had})$, where $\tilde{N}/\varepsilon$ is the efficiency corrected number of resonance decays.
Here $\Gamma_{\rm had}$ includes all decay modes of the resonances other than $e^+e^-$, $\mu^+\mu^-$, and $\tau^+\tau^-$.
Then, assuming lepton universality
${\cal B}_{\mu\mu} = \Gamma_{\mu\mu}/\Gamma = {\tilde{\cal{B}}}_{\mu\mu}/(1+3{\tilde{\cal{B}}}_{\mu\mu})$.

The off-resonance data are used to subtract the dominant background from the on-resonance data due to non-resonant (continuum) production of $\mu^+\mu^-$ and hadrons via $e^+ e^- \to \mu^+\mu^-$ and $e^+ e^- \to q\bar{q}$ ($q=u,d,c,s$): $\tilde{N} = \tilde{N}_{on} - S \tilde{N}_{off}$. 
The factor $S$ scales the luminosity of the off-resonance sample to that of the on-resonance sample taking into account the $1/s$ dependence of the cross section.
Backgrounds from other non-resonant sources are negligible after the off-resonance subtraction.

The remaining backgrounds (to $\mu^+\mu^-$) are mainly from cosmic rays, and at the $\Upsilon$(2S) and $\Upsilon$(3S) from cascade decays to a lower $\Upsilon$ state, which decays to $\mu^+\mu^-$ while the accompanying particles escape detection. 
The background from $\Upsilon \to \tau^+\tau^-$ is also significant in the hadron measurement.

\section{Event selection}

We select $\mu^+\mu^-$ events by requiring exactly two oppositely charged tracks, each with momentum $0.7 < p/E_{\rm beam} < 1.15$, with polar angle $| \cos\theta | < 0.8$, and with opening angle of the tracks greater than 170$^{\circ}$.
Muon identification requires each track to deposit $0.1-0.6$ GeV in the electromagnetic calorimeter, characteristic of a minimum ionizing particle, and at least one track to penetrate deeper than five interaction lengths into the muon chambers.

We control the cosmic-ray background using the track impact parameters with respect to the $e^+e^-$ interaction point (beam spot).
The average distance of the tracks from the beam spot is especially useful to suppress cosmic rays and to estimate the remaining background in our data samples.
We correct the number of $\mu^+\mu^-$ events observed in the on-resonance and off-resonance samples individually for this background.

We also require fewer than two extra showers with more than 50 MeV (100 MeV) energy in the barrel (endcap) section of the calorimeter in order to reduce the indirect $\mu^+\mu^-$ production at the $\Upsilon$(2S) and $\Upsilon$(3S) when these cascade down via $\pi^0\pi^0$ or $\gamma\gamma$ emission to a lower $\Upsilon$ resonance which decays to $\mu^+\mu^-$.
The residual cascade background is estimated to be ($2.9\pm 1.5$)\% and ($2.2\pm 0.7$)\% for $\Upsilon$(2S) and $\Upsilon$(3S), respectively.

\begin{figure}
\begin{center}
\includegraphics*[width=3.4in]{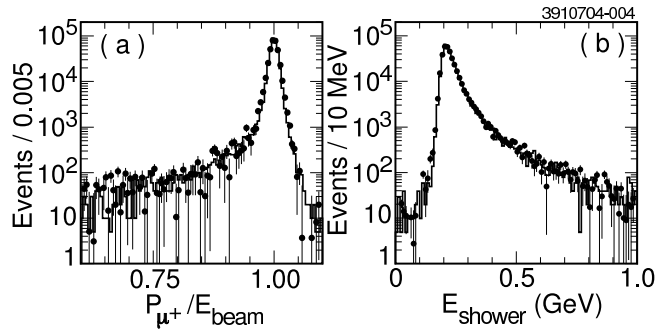}
\end{center}
\caption{Distribution of the scaled momentum (left) and the shower energy (right) of the $\mu^+$ candidates from $\Upsilon$(1S) decays after off-resonance subtraction (points) and from resonance Monte Carlo simulations (histogram). The vertical scale is logarithmic.}
\label{fig:data_mc}
\end{figure}

The overall selection efficiency for $\Upsilon \to \mu^+\mu^-$ decays is ($65.2 \pm 1.2$)\% from a Monte Carlo simulation of the detector response (Figure \ref{fig:data_mc}).
The relative systematic uncertainty in the efficiency is $1.8$\% which is dominated by the uncertainty in detector simulation ($1.7$\%).

When selecting hadronic events we suppress QED backgrounds ($e^+e^- \to e^+e^-/\mu^+\mu^-/\gamma\gamma$) as well as events from beam-gas and beam-wall interactions by a combination of cuts on the number of tracks, total visible energy, calorimeter energy and the vertex position. 

The overall selection efficiency for hadronic resonance decays varies between 96-98\% (largest for the 1S and smallest for the 2S) depending on the relative rate of cascade decays that can produce a stiff $e^+e^-$ or $\mu^+\mu^-$ in the final state.
The relative systematic uncertainty in the selection efficiency is between $1.3-1.6$\% and dominated by the uncertainties in the simulation of hadronization and detector modeling.

Background from $\Upsilon \to \tau^+\tau^-$ decay is estimated to be between $0.4-0.7$\%.

\section{Results}

\begin{figure}
\includegraphics*[width=3.4in]{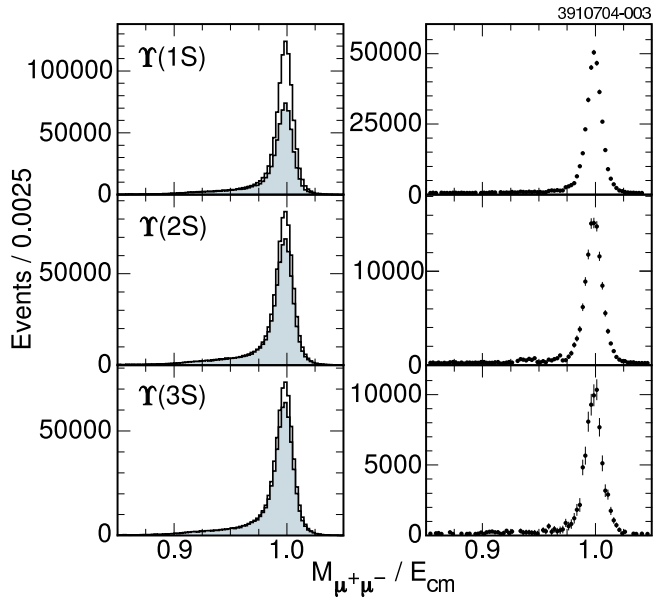}\hspace{2pc}%
\begin{minipage}[b]{2.5in}\caption{Muon pair invariant mass distributions in on-resonance (empty) and scaled off-resonance (shaded) data on the left and the difference between these two distributions on the right.}
\end{minipage}
\label{fig:muons}
\end{figure}

Table~\ref{tab:result} presents the numerical results and comparison with the current world average values of ${\cal B}_{\mu\mu}$ \cite{PDG}.
The invariant mass distribution of the $\mu^+\mu^-$ candidates in the on-resonance and off-resonance samples and after off-resonance subtraction are shown in Fig.~\ref{fig:muons}.

\begin{table}
  \centering
  \caption{Number of resonance decays to $\mu^+\mu^-$ and hadrons ($\tilde{N}$), selection efficiencies ($\varepsilon$), and the muonic branching fractions after correcting for interference. The current world averages \cite{PDG} are also listed in the last line for comparison.}\label{tab:result}
\bigskip
\begin{tabular}[c]{lccc}
  \hline
  \hline
    & $\Upsilon$(1S) & $\Upsilon$(2S) & $\Upsilon$(3S) \\
  \hline
   $\tilde{N}_{\mu\mu}$ ($10^3$) & $344.9 \pm 2.5 \pm 0.3$ & $119.6 \pm 1.8 \pm 1.9$ & $81.2 \pm 2.7 \pm 0.7$ \\
   $\varepsilon_{\mu\mu}$ & $0.652\pm0.002 \pm 0.012$ & $0.652\pm0.002 \pm 0.012$ & $0.652\pm0.002 \pm 0.012$ \\
   $\tilde{N}_{\rm had}$ ($10^6$) & $18.96 \pm 0.01 \pm 0.04$ & $7.84 \pm 0.01 \pm 0.02$ & $4.64 \pm 0.01 \pm 0.02$ \\
   $\varepsilon_{\rm had}$ & $0.979 \pm 0.001 \pm 0.016$ & $0.965 \pm 0.001 \pm 0.013$ & $0.975 \pm 0.001 \pm 0.014$ \\
  \hline
   ${\cal B}_{\mu\mu}$ (\%) & $2.49 \pm 0.02 \pm 0.07$ & $2.03 \pm 0.03 \pm 0.08$ & $2.39 \pm 0.07 \pm 0.10$ \\
  \hline
   ${\cal B}_{\mu\mu}$ (\%) PDG & $2.48 \pm 0.06$ & $1.31 \pm 0.21$ & $1.81 \pm 0.17$ \\
  \hline
  \hline
\end{tabular}
\end{table}

We have corrected ${\cal B}_{\mu\mu}$ for the interference between resonant and non-resonant production of $\mu^+ \mu^-$ (and hadron) final states by reducing the measured branching fractions by a relative $2-4$\%.

The total fractional systematic uncertainties in ${\cal B}_{\mu\mu}$ are 2.7\% (1S), 3.7\% (2S), and 4.1\% (3S), respectively.
They include the uncertainties in the number of resonance decays due to background subtraction and in the selection efficiencies due to detector modeling, trigger efficiencies and Monte Carlo statistics.
Uncertainties in the interference calculation and variations in the center-of-mass energy contribute an additional 1\% to the fractional uncertainty of ${\cal B}_{\mu\mu}$.
The dominant source of the systematic uncertainty in the cases of $\Upsilon$(2S) and $\Upsilon$(3S) is due to the uncertainty in the scale factor between the on-resonance and off-resonance data.

The result for the $\Upsilon$(1S) is in very good agreement with the current world average, while our $\Upsilon$(2S) and $\Upsilon$(3S) results are about $3\sigma$ larger than the world averages.

The improved muonic branching fractions, combined with the current values of $\Gamma_{ee}\Gamma_{\rm had}/\Gamma$ \cite{PDG} lead to the following new values for the total decay widths of the three narrow $\Upsilon$ resonances: $\Gamma (1S) = (52.8 \pm 1.8)$ keV, $\Gamma (2S) = (29.0 \pm 1.6)$ keV, and $\Gamma (3S) = (20.3 \pm 2.1)$ keV.

\section{Conclusion}

CLEO has measured the muonic branching fraction of the narrow $\Upsilon$ resonances below the open-beauty threshold with 2.8\%, 4.0\%, and 5.1\% relative uncertainty.
The obtained branching fractions for the $\Upsilon$(2S) and $\Upsilon$(3S) resonances are significantly larger than prior measurements and the current world average values, resulting in narrower total decay widths.
The new branching fractions, particularly ${\cal B}_{\mu\mu}(2S)$, affect the measured rates of other transitions leading to the $\Upsilon$ resonances and observed by the subsequent decay $\Upsilon \to \mu^+\mu^-$.
The new total widths of the $\Upsilon$(2S) and $\Upsilon$(3S), have a significant impact on the comparison between theoretical and experimental values of hadronic and radiative widths of these resonances.

\ack

We gratefully acknowledge the effort of the CESR staff 
in providing us with excellent luminosity and running conditions.
This work was supported by the National Science Foundation
and the U.S. Department of Energy.

\smallskip


\medskip
\section*{References}
\begin{thebibliography}{6}

\bibitem{Yreview} Besson D and Skwarnicki T 1993 {\it Ann. Rev. Nucl. Part. Sci.} {\bf 43} 333 and references therein.

\bibitem{LQCD} Thacker B A and Lepage G P 1991 {\it Phys. Rev.} D {\bf 43} 
196
\nonum Kronfeld A S and Mackenzie P B 1993 {\it Ann. Rev. Nucl. Part. Sci.} 
{\bf 43} 793
\nonum Davies C T H {\it et al}. 1998 {\it Phys. Rev.} D {\bf 58} 054505
\nonum Davies C T H {\it et al}. 2004 {\it Phys. Rev. Lett.} {\bf 92} 022001

\bibitem{PDG} Eidelman S {\it et al}. (Particle Data Group) 2004 {\it Phys. Lett.} B {\bf 592} 1

\bibitem{Higgs} Sanchis-Lozano M A 2004 {\it Int.J. Mod. Phys.} A {\bf 19} 2183 ({\it Preprint} hep-ph/0307313)
\nonum Sanchis-Lozano M A 2004 {\it Preprint} hep-ph/0401031

\bibitem{early_exp} Andrews D {\it et al}. (CLEO Collaboration) 1983 {\it Phys. Rev. Lett.} {\bf 50} 807
\nonum Haas P {\it et al}. (CLEO Collaboration) 1984 {\it Phys. Rev.} D {\bf 30} 1996
\nonum Albrecht H {\it et al}. (ARGUS Collaboration) 1985 {\it Z. Phys.} C {\bf 28}
\nonum Kaarsberg T M {\it et al}. (CUSB Collaboration) 1989 {\it Phys. Rev. Lett.} {\bf 62} 2077
\nonum Chen W-Y {\it et al}. (CLEO Collaboration) 1989 {\it Phys. Rev.} D {\bf 39} 3528
\nonum Kobel M {\it et al}. (Crystal Ball Collaboration) 1992 {\it Z. Phys.} C {\bf 53} 193

\bibitem{CLEO_new} Adams G S {\it et al}. (CLEO Collaboration) 2004 {\it Preprint} hep-ex/0409027 (accepted by Phys. Rev. Lett.)

\end{thebibliography}
\end{document}